\documentclass[aps,preprint,nofootinbib,floatfix,superscriptaddress,prd]{revtex4-2}
\usepackage{graphicx,color}
\usepackage{booktabs}
\usepackage{amsmath,amssymb,mathtools}
\usepackage{dsfont}
\usepackage[pdfusetitle,pdfencoding=auto,colorlinks]{hyperref}
\hypersetup{
    linkcolor=magenta,
    citecolor=magenta,
    urlcolor=blue,
    }

\newcommand{\be}{\begin{eqnarray}}
\newcommand{\bb}{\bibitem}
\newcommand{\ee}{\end{eqnarray}}

\newcommand{\fig}{\begin{figure}}
\newcommand{\ef}{\end{figure}}
\newcommand{\bc}{\begin{center}}
\newcommand{\ec}{\end{center}}

\newcommand{\nm}{\mbox}

\newcommand{\bn}{\begin{enumerate}}
\newcommand{\en}{\end{enumerate}}

\newcommand{\bz}{\begin{itemize}}
\newcommand{\ez}{\end{itemize}}

\newcommand{\ba}{\begin{array}} % \lf \ba {ccc}  a & b & c \ea \ri
\newcommand{\ea}{\end{array}}

\newcommand{\bt}{\begin{tabular}}
\newcommand{\et}{\end{tabular}}

\newcommand{\bd}{\begin{displaymath}}
\newcommand{\ed}{\end{displaymath}}
\newcommand{\nn}{\nonumber}
\newcommand{\ben}{\begin{eqnarray*}}
\newcommand{\een}{\end{eqnarray*}}

\newcommand{\bq}{\begin{quote}}
\newcommand{\eq}{\end{quote}}

\newcommand{\gsim}{\gtrsim}
\newcommand{\lsim}{\lesssim}
\def\bea{\begin{eqnarray}}
\def\eea{\end{eqnarray}}
\def\nn{\nonumber}

\renewcommand\epsilon{\varepsilon}

\def\lsim{\mathrel{\raise.3ex\hbox{$<$\kern-.75em\lower1ex\hbox{$\sim$}}} }
\def\gsim{\mathrel{\raise.3ex\hbox{$>$\kern-.75em\lower1ex\hbox{$\sim$}}} }

\begin{document}

\title{Implications of the new CDF-II $W$-boson mass on two-Higgs-doublet models}
\author{Yang Hwan Ahn}
\email{axionahn@naver.com}
\author{Sin Kyu Kang}
\email{skkang@seoultech.ac.kr}
\author{Raymundo Ramos}
\email{rayramosang@gmail.com}
\affiliation{%
School of Liberal Arts, Seoul National University of Science and Technology,
232 Gongneung-ro, Nowon-gu, Seoul 01811, Korea
}

\date{\today}

\begin{abstract}
We present the implications of the recent measurement of $W$ boson at CDF II on the two-Higgs-doublet model (2HDM).
In the analysis, we impose theoretical bounds such as vacuum stability and perturbative unitarity, and several experimental constraints.
In addition, we take into account the measurement of $\sin^2\theta_W(m_Z)_{\rm \overline{MS}}$ on top of the CDF $W$-boson mass to investigate how  the $S$ and $T$ parameters are determined. 
We explore two possible scenarios depending on whether the Higgs boson observed at the LHC is the lighter or heavier of $CP$-even neutral Higgs bosons 
for 2HDM type I and II\@.
Using the results, we show how the parameter space is constrained, and compare it with the one based on the PDG average of $m_W$.
Furthermore, we explore phenomenological consequences of electroweak precision observables that can be affected by $m_W$ within the predictions of the 2HDM, and the reduction in parameter space
expected from future measurements
% on electroweak precision observables
at the Future Circular Lepton Collider.

\end{abstract}

% =======================

\maketitle
\newpage

\section{Introduction}
Very recently, CDF announced a measurement of the $W$ boson mass~\cite{cdf}
\begin{equation}
    m_{W}^{\rm CDF}=80.4335\pm 0.0094\mbox{ GeV}. \label{cdf}
\end{equation}
This result represents two intriguing points.
One is that it is an unprecedented, highly precise measurement of $m_W$, and the other is that it is in about $7 \sigma$ tension with the prediction of the standard model (SM), which is $m_W^{\rm SM}=80.379\pm 0.006$~GeV~\cite{Awramik:2003rn}.
Although the CDF result of $m_W^{\rm CDF}$ also shows a significant shift  compared to the PDG average of the LEP~\cite{ALEPH:2013dgf},
ATLAS~\cite{ATLAS:2017rzl} and the previous Tevatron~\cite{CDF:2013dpa} results yielding $m_W^{\rm PDG}=80.379\pm 0.012$~GeV~\cite{ParticleDataGroup:2020ssz} as well as the result from LHCb leading to
$m_W^{\rm LHCb}=80.354\pm 0.031$~GeV~\cite{lhcb},
 it may serve as a hint of new physics beyond the SM~\cite{skk,wmh}.
Under the assumption that the CDF measurement will be confirmed in the foreseeable future, it would deserve to  explore its phenomenological implications.

The purpose of this work is to examine the implication of  the recent measurement of  the $W$ boson mass at CDF II on two-Higgs-doublet models (2HDMs).
Recently, this possibility has been explored with different approaches in other works~\cite{Fan:2022dck,Zhu:2022scj,Song:2022xts,Bahl:2022xzi,Babu:2022pdn,Heo:2022dey}
and comprehensive analyses of the parameter space and other phenomenology have been performed not long ago (see, for example, Refs.~\cite{Atkinson:2021eox,Atkinson:2022pcn,Botella:2018gzy,Broggio:2014mna}).
The deviation of $m_W^{\rm CDF}$ from its SM prediction can be parametrized in terms of the so-called Peskin-Takeuchi parameters, $S$ and $T$,
which represent the contributions of new physics.
The quantum corrections mediated by new scalar fields contribute to $S$ and $T$.
It is common belief that the exquisite precision achieved in the measurements of $m_Z$, $m_W$, and $\sin^2 \theta_W$ makes it possible to explore the existence of new physics beyond the SM~\cite{ST}.
In this regard, another important parameter to test the SM is the so-called weak mixing angle parameter, $\sin^2\theta_W$~\cite{ST}.
Instead of on-shell definition of $\sin^2\theta_W$, it is more general to take it into account by employing the more theoretically motivated
$\overline{\rm MS}$ (modified minimal subtraction) prescription.
All $Z^{0}$ pole measurements of $\sin^2\theta_W(m_Z)_{\overline{\rm MS}}$ are usually averaged out to give
\begin{equation}
    \sin^2 \theta_W(m_Z)_{\overline{\rm MS}}^{\rm ave}=0.23124\pm 0.00006. \label{sin2thetaw}
\end{equation}
Equation~(\ref{sin2thetaw}) represents that the average of all $Z^{0}$ pole measurements is in consistent with the SM\@.
But, the contributions of new physics to $\sin^2\theta_W(m_Z)_{\overline{\rm MS}}$ can also be parametrized in terms of $S$ and $T$ parameters.
Thus, the deviation of $m_W$ from its SM prediction may, in general, affect the prediction of $\sin^2\theta_W(m_Z)_{\overline{\rm MS}}$.
In this work, we will first examine whether there exist nontrivial $S$ and $T$ parameters accommodating both $m_W^{\rm CDF}$ and $\sin^2 \theta_W(m_Z)_{\overline{\rm MS}}^{\rm ave}$, and compare them with those obtained from the global fit of Ref.~\cite{Lu:2022bgw}.
Using the allowed regions of the parameter $S$ and $T$ from two observables $m_W$ and $\sin^2\theta_W(m_Z)_{\overline{\mbox{MS}}}$, we will estimate the allowed regions of the masses of new scalar bosons and mixing angles in 2HDMs\@.
 To see the impact of the recent CDF $W$-boson mass on the parameter scan, we compare the results based on $m_W^{\rm CDF}$ with
those based on $m_W^{\rm PDG}$.
%
% estimate how the deviation of $m_W^{\rm CDF}$ can be accommodated in the frame work 2HDM by scanning masses of new scalars, and then try %to obtain allowed regions of the parameter space.
%Next, we will , and then confront them with the parameter space obtained from the first stage.
In the analysis, we impose the theoretical conditions such as  the vacuum stability,
perturbativity~\cite{2HDMvacuum, Ferreira} and  unitarity~\cite{unitarity,Eriksson:2009ws}, and experimental constraints to constrain the masses of Higgs fields and mixing parameters.
Since the 125.1 GeV Higgs boson observed at the LHC
can be either the lighter or heavier $CP$-even neutral scalar boson in the 2HDM,
we divide our study in two scenarios,
labeling as scenario 1 the case when the lighter Higgs mass,
$M_h$, corresponds to the 125.1~GeV scalar;
and scenario 2 to whenever the heavier Higgs takes that place.
In addition, we study the phenomenological consequences of other observables that can be affected by $m_W$
within the predictions of the 2HDM, most notably, the decay width of the $Z$ boson.
We expand on this idea by considering measurements of electroweak precision observables
in the proposed Future Circular Lepton Collider (FCC-ee).
The FCC-ee is the first step in the Future Circular Collider integrated program~\cite{Benedikt:2022wvj},
and will consist of a 100 km underground circular machine that will take data over 15 years.
Its implementation will follow an staged approach focusing on electroweak, flavor, Higgs, and top physics.
For our analysis, we will focus on the electroweak precision observables prospects where the FCC-ee is expected to reduce the current uncertainties around 500 times~\cite{Blondel:2021ema}
and see how they can distinguish both cases of
$m_W$.

%The theoretical conditions taken into account are the vacuum stability, perturbativity and unitarity which are required to be satisfied up to a cut-off scale. Then one can obtain constraints on the couplings of the Higgs potential
%in 2HDM, which in turn lead to bounds on the masses of scalar bosons as well as mixing parameters.

The rest of the paper is laid out as follows:
In Sec.~\ref{sec:the2hdm}, we give a brief summary of 2HDMs and  present how the observables $m_W$ and $\sin^2\theta_W$ can be parametrized in terms of $S$ and $T$ parameters. The allowed regions of $S$ and $T$ are obtained by imposing the recent CDF $m_W$ and the PDG result of $\sin^2\theta(m_Z)_{\overline{\rm MS}}$, and compare them with those obtained from the global fit of Ref.~\cite{Lu:2022bgw}.
We also discuss on the various constraints from theoretical conditions and experimental results, which will be imposed in this analysis.
In Sec.~\ref{sec:numresults}, we show our numerical results and discuss the implications of CDF's measurement of $m_W$ by incorporating the PDG result of $\sin^2\theta(m_Z)_{\overline{\rm MS}}^{\rm ave}$ in the scenarios mentioned above for 2HDM types I and II\@.
Finally, in Sec.~\ref{sec:conclusion} we discuss the most relevant details of this work and conclude.

\section{Two-Higgs-doublet model}
\label{sec:the2hdm}

\subsection{The setup}

Taking  $\Phi_1$ and $\Phi_2$ are two complex $SU(2)_L$ Higgs doublet  fields with  $Y=1$, 
the renormalizable gauge  invariant  scalar potential of 2HDM with softly broken $Z_2$ symmetry  under which $\Phi_1 \rightarrow \Phi_1$ and $\Phi_2 \rightarrow - \Phi_2$  is written as~\cite{2HDM}
\begin{align}
    V= {}& m^2 _{11} \Phi^\dagger _1 \Phi_1 + m^2 _{22} \Phi^\dagger _2 \Phi_2 - (m^2 _{12} \Phi^\dagger _1 \Phi_2 + \nm{h.c.}) 
        + \frac{1}{2}\lambda_1 (\Phi^\dagger _1 \Phi_1)^2 + \frac{1}{2}\lambda_2 (\Phi^\dagger _2 \Phi_2 )^2 \nn \\
    & + \lambda_3 (\Phi^\dagger _1 \Phi_1) (\Phi^\dagger _2 \Phi_2)  + \lambda_4 (\Phi^\dagger_1 \Phi_2 ) (\Phi^\dagger _2 \Phi_1)  + \Big \{ \frac{1}{2} \lambda_5 (\Phi^\dagger _1 \Phi_2 )^2
    + \nm{h.c.} \Big \}.  \label{2hd pot}
\end{align}
We note that all the parameters in Eq.~(\ref{2hd pot}) to be real and the squared mass of pseudoscalar $m^2 _A$ to be greater than  $ |\lambda_5| v^2$ so as to keep $CP$ symmetry in the scalar potential.\footnote{ The other terms generally allowed in the scalar potential  are ignored to keep $CP$ symmetry.}
As one can easily check, the dangerous flavor changing neutral currents are absent in the form given by Eq.~(\ref{2hd pot}) even if  nonzero $m^2_{12}$ softly breaking the $Z_2$ symmetry is allowed.

Yukawa interactions of $h$ and $H$ are parametrized by
\begin{equation}
{\cal L}_{\rm yuk} = - \sum_{f=u,d,l}\frac{m_f}{v} \Big( \hat{y}^h_f \bar{f}f h+\hat{y}^H_f \bar{f}f H\Big),
\end{equation}
where the effective couplings of $\hat{y}^{h,H}_f $ are referred to~\cite{sk1}.
Depending on how to couple the Higgs doublets to the fermions, 2HDMs are classified into four types~\cite{Branco}. Among them, the Yukawa couplings of 2HDM type II arises in the minimal supersymmetric standard model which is one of the most promising candidates for the new physics model beyond the SM\@.
In this work, we consider 2HDM type I and type II, and present how the allowed regions of the masses of new scalars and mixing angles are different from each other.
%In this paper, we focus on the type-II 2HDM, in which the one Higgs doublet $\Phi_1$ couples only to the down type quarks and the charged leptons while the another Higgs doublet $\Phi_2$ couples only to the up-type quarks.
 
The spontaneous breaking of electroweak symmetry triggers the generation of the vacuum expectation values of the Higgs fields as follows:
\begin{equation}
\langle\Phi_1\rangle = \frac{1}{\sqrt{2}} \left ( \ba{c} 0 \\ v_1 \ea \right ), \qquad \langle\Phi_2\rangle = \frac{1}{\sqrt{2}} \left ( \ba{c} 0 \\ v_2 \ea \right ),
\end{equation}
where
$ v^2 \equiv v^2 _1 + v^2 _2 =  (246 ~\mbox{GeV})^2$ with $v_2/v_1 = \tan\beta $, and
$v_1$ and $v_2$ are taken positive, so that $0\leq \beta \leq \pi/2$ is allowed.
Then, the fluctuation fields around $v_1$ and $v_2$ become
\begin{equation}
\Phi_1= \left ( \ba{c} \phi_1^{+} \\ \frac{ v_1+\rho_1+i \eta_1}{\sqrt{2}}   \ea \right ), \qquad \Phi_2 =  \left ( \ba{c} \phi_2^{+} \\ \frac{ v_2+\rho_2+i \eta_2}{\sqrt{2}} \ea \right ).
\end{equation}
Among the 8 degrees of freedom, three are eaten by the gauge bosons and the remaining five become physical Higgs particles in 2HDM: two $CP$-even neutral Higgses $h$ and $H$ ($M_h \leq M_H$), a $CP$-odd neutral Higgs $A$ and a charged Higgs pair ($H^\pm$).
The neutral scalars are given by
\begin{align}
    h & =\sqrt{2}(\eta_2 \sin \alpha - \eta_1 \cos\beta), \nonumber \\
    H & =-\sqrt{2}(\eta_2 \cos\alpha + \eta_1 \sin\beta), \\
    A & =\sqrt{2}(\rho_2 \sin\beta - \rho_1 \cos\beta)\nonumber
\end{align}
Following Ref.~\cite{2HDM},
the squared masses for the $CP$-odd and charged Higgs states are calculated to be
\begin{equation}
M^2 _A = \frac{m^2 _{12}}{s_\beta c_\beta} -\lambda_5 v^2 ,  \quad
M^2 _{H^\pm} =  M^2 _A + \frac{1}{2} v^2 (\lambda_5 - \lambda_4),
\label{cp odd charged masses}
\end{equation}
and the squared masses for  neutral Higgs ($M_H \geq M_h$) are given by~\cite{2HDM}
\begin{equation}
M^2 _{H, h} = \frac{1}{2}\Big  [ P+Q \pm \sqrt{(P - Q )^2 + 4 R^2 }~ \Big ], \label{cp even mass 2}
\end{equation}
where
$P=\lambda_1 v^2_1+m^2_{12} t_\beta$, $Q=\lambda_2 v^2_2+m^2 _{12}/ t_\beta$ and $R=(\lambda_3+\lambda_4+\lambda_5)v_1v_2 -m^2 _{12}$ with $s_\beta = \sin\beta$, $c_\beta = \cos\beta$, and $t_\beta=\tan\beta$.
While $h$ becomes the SM-like Higgs boson  for $\sin(\beta-\alpha)=1$, $H$ does so for $\cos(\beta-\alpha)=1$.
In this work, we examine the implications of the recent CDF $W$-boson mass in both cases separately.
%The couplings of the two neutral CP even Higgs bosons to fermions and bosons relative to the SM couplings in type II 2HDM are shown in Table \ref{tab1}.
%\begin{table}
%\caption{\label{tab1} Neutral Higgs couplings relative to the SM couplings in Type II 2HDM. $D,L,U,W,Z$ and $A$ stand for down-type quarks, charged leptons, up-type quarks, two weak gauge bosons and CP odd Higgs, respectively.}
%
%\begin{tabular}{|c|c|c|}
%\hline \hline
%& Light Higgs ($h$) & Heavy Higgs ($H$) \\
%\hline
%D,L &  $- \frac{\sin\alpha}{\cos\beta}$ & $\frac{\cos\alpha}{\cos\beta}$ \\
%U & $\frac{\cos\alpha}{\sin\beta}$ & $\frac{\sin\alpha}{\sin\beta}$ \\
%W or Z & $\sin(\beta -\alpha)$ & $\cos(\beta -\alpha)$ \\
%AZ & $- \cos(\beta -\alpha)$ & $-\sin(\beta -\alpha)$ \\
%\hline \hline
%\end{tabular}
%
%\end{table}

\subsection{$W$-boson mass and $\sin^2 \theta_W (m_Z)_{\overline{\rm MS}}$}
 The contribution of new scalar fields to $T$ and $S$ parameters are given by~\cite{Toussaint,kanemura,Baak,Neil}
\begin{align}
T&= \frac{\sqrt{2}G_F}{16\pi^2 \alpha_{EM}} \Big\{
    -F'(M_A, M_H^\pm) + \sin^2(\beta - \alpha)\Big[F'(M_H, M_A) - F'(M_H, M_{H^\pm})\Big]\nn\\
 & \qquad \qquad \qquad + \cos^2(\beta - \alpha)\Big[F'(M_h, M_A) - F'(M_h, M_{H^\pm})\Big]
    \Big\}, \label{T parameter} \\
S &= - \frac{1}{4\pi} \Big[ F(M_{H^\pm}, M_{H^\pm}) - \sin^2 (\beta -\alpha) F (M_H, M_A) - \cos^2 (\beta -\alpha) F(M_h . M_A) \Big],  \label{S parameter}
\end{align}
where 
%$\Delta \rho^{\rm new} _0 = \Delta \rho^{\rm 2HDM} - \Delta \rho^{\rm SM}$ and the formulae for $\Delta \rho^{\rm 2HDM}$ as well as $\Delta \rho^{\rm SM}$ are given in \cite{hunter, cheung, Chankowski}, and 
the functions $F$ and $F'$ are given by~\cite{hunter, cheung, Chankowski,Toussaint,kanemura,Baak}
\begin{align}
    F(x,y)  = {}& - \frac{1}{3} \Big[\frac{4}{3} - \frac{x^2 \ln x^2 - y^2 \ln y^2}{x^2 - y^2}
 - \frac{x^2 + y^2}{(x^2 - y^2)^2} \Big ( 1+\frac{x^2 +y^2}{2} - \frac{x^2 y^2}{x^2 - y^2}\ln\frac{x^2}{y^2}\Big ) \Big],\\
    F'(x,y) = {}& \frac{x^2 + y^2}{2} - \frac{x^2y^2}{x^2 - y^2}\ln\frac{x^2}{y^2}.
\end{align}

Employing the precise measurements of QED coupling $\alpha$,  $G_F$ and $m_Z$ accompanied by
 $m_t$ and $ M_{h_{\rm SM}} = 125.1~ \mbox{GeV}$, and allowing for loop effects mediated by  heavy new particles
via $S$ and $T$ parameters, 
we can recast the expressions for the predictions of $m_W$ and $\sin^2 \theta_W (m_Z)_{\overline{\rm MS}}$ as follows~\cite{ST}
\begin{align}
  m_W & = 80.357 ~\mbox{GeV}~(1-0.0036~ S+ 0.0056~ T), \label{st-values-mw}\\ 
\sin^2 \theta_W (m_Z)_{\overline{\rm MS}} & = 0.23124 ~(1+0.0157~ S - 0.0112~ T). \label{st-values}
\end{align}
Plugging the experimental values for $m_W$, from Eq.~(\ref{cdf}), and $\sin^2 \theta_W (m_Z)_{\overline{\rm MS}}$, from Eq.~(\ref{sin2thetaw}),
into Eqs.~\eqref{st-values-mw} and~(\ref{st-values}) we obtain the allowed regions of $S$ and $T$ parameters as follows;
\begin{align}
T &= 0.3\pm 0.062, \nn \\
S &= 0.2\pm 0.08. \label{ew-parameter}
\end{align}
%We impose the conditions Eq.(\ref{ew-parameter}) in our numerical analysis.
Those results may indicate that the contributions of new physics are prominent in $m_W$ while
they are canceled in $\sin^2\theta_W(m_Z)_{\overline{\rm MS}}$.
Note that these values are in agreement with the values obtained in Ref.~\cite{Lu:2022bgw}.
To be precise, the value of $S$ is shifted $\sim$25\% in our work with the same error of roughly 50\%.
In the case of $T$ the shift is only  $\sim$10\% and the errors (roughly 20\%) are almost identical.
In our numerical analysis, we study how the masses of new scalar fields and mixing angles in 2HDM can be constrained by imposing the results of Eq.~(\ref{ew-parameter}). Thus, our results are more affected by the CDF $m_W$ value than from directly using the values from the global fit.

\subsection{The bounds}
\label{sec:thebounds}

The vacuum stability of the scalar potential, Eq.~(\ref{2hd pot}), is guaranteed only if  the following conditions are satisfied~\cite{Ferreira, 2HDM}
\begin{equation}
\lambda_{1,2} > 0, \quad  \lambda_3 > -\sqrt{\lambda_1 \lambda_2}, \quad  \lambda_3 + \lambda_4 - |\lambda_5| > - \sqrt{\lambda_1 \lambda_2}.  \label{stability}
\end{equation}
%Since radiative corrections lead to the modification of the couplings in the scalar potential,
%it is required that  the stability conditions (\ref{stability}) are valid for high energy up to cut-off scale $\Lambda$.
The stability conditions of Eq.~(\ref{stability}) give rise to lower bounds on
the couplings $\lambda_i$~\cite{2HDM}, which in turn lead to bounds on the masses of the physical Higgs fields.
In addition,  we require that  the quartic couplings $\lambda_i $ in the scalar potential is perturbative and unitarity conditions~\cite{unitarity} are satisfied.
% even at high scale up to the cut-off scale. 
Those theoretical conditions can constrain not only the masses of the Higgs fields but also mixing parameters $\tan\beta$ and $\alpha$.
%In our numerical analysis, we used RG equations for the parameters 
%$m_{ii}^2, \lambda_{i}$ , gauge couplings $g_{i}$ and Yukawa couplings presented in ref.\cite{RGrunning}.
%In particular, we take the top quark pole mass and QCD coupling constant at Z boson mass scale ($\alpha_s(M_Z)$)  to be 172 GeV and 0.1185, respectively.
%

On the other hand, we can consider the experimental constraints.
Since the Higgs-strahlung at the LEP is one of the most direct channels to probe a light Higgs boson $h$ with mass below 120 GeV, we use the strongest
upper bound on the event rate of $e^+e^- \rightarrow Zh \rightarrow Z jj$~\cite{LEP,LEP2} to constrain  the mass and  mixing parameters.
We also consider  the constraints coming from the Higgs pair production process, $e^+ e^- \rightarrow hA  \rightarrow b\bar{b}b\bar{b}$ when the mass parameters are kinematically allowed~\cite{LEP2}.
The experimental lower bound on charged Higgs masses is 79.3~GeV~\cite{ALEPH}.
 The nonobservation of $Z\rightarrow h A$ in the LEP experiment gives rise to the condition that  $M_h + M_A > M_Z$ are kinematically allowed~\cite{zdecay}.
%In addition,  when $M_h \lesssim 115$ GeV, non-observation of the Higgsstrahlung process $e^+ e^- \rightarrow hZ \rightarrow b\bar{b}Z$  at the LEP constrains the parameter space of $\sin^2(\beta-\alpha)\times Br(h\rightarrow b\bar{b})$ and $M_h$\footnotemark
%\addtocounter{footnote}{0}
%\footnotetext{The parameter $\xi^2$ introduced in \cite{LEP} is equivalent to
%$\sin^2(\beta - \alpha) $ in our model.}.
We include the upper bounds on ${\rm Br}(t \rightarrow H^+ b) \times {\rm Br}(H^+ \rightarrow \tau^+ \nu_{\tau})$ coming from the LHC search for $H^\pm$
through the channel $pp \rightarrow t\bar{t}\rightarrow b\bar{b}H^{\pm}W^{\mp}$ followed by $H^{\pm}\rightarrow \tau^{\pm}\nu$~\cite{charged1,charged2,charged3}.
The results from the LHC Higgs signal strengths at 7 and 8~TeV are taken into account.
I%n scenario 2, the pseudo-scalar $A$ is the only heavy neutral scalar and no significant excess in the heavy neutral Higgs search can provide the exlucsion limit.
%For the bounds associated with $A$, we consider $gg\rightarrow A \rightarrow \gamma \gamma$\cite{lhca1,lhca2}, $gg\rightarrow A \rightarrow \tau^+ \tau^-$, and $b\bar{b}\rightarrow A \rightarrow \tau^+ \tau^-$.

In addition to the above constraints explained, we take into account the measurement of  $R_b \equiv \Gamma (Z \rightarrow b \bar{b})/\Gamma(Z \rightarrow \mbox{hadrons})$~\cite{ParticleDataGroup:2020ssz}.
The updated SM prediction of $\mbox{B}_{\rm SM}(\bar{B}\rightarrow X_s \gamma)$~\cite{bsg} and the Belle experiment~\cite{bsg2}
give severe bound on $M_{H^{\pm}}$ in the type II;
with the lower bound at 95\% C.L. for $M_{H^{\pm}}$ in the range 570--800~GeV~\cite{Misiak:2017bgg}.
% $M_{H^{\pm}}>570$ GeV for $t_{\beta}\gtrsim 2$ at $95 \%$ C.L.
%the experimental results of the process $b\rightarrow s \gamma$ \cite{bsg,bsg2}, which lead to the constraints on the $M_{H^\pm} - \tan \beta$ plane.
% In the Type-II 2HDM,  it is known that $R_b$ yields the strictest bound on the $M_{H^\pm} - \tan \beta$ plane in the small $\tan\beta$ region  \cite{rb,d2hdm}. 
The measurements of $B-\bar{B}$ mixing also lead to the constraints on  the $M_{H^\pm} - \tan\beta$ plane but  less severe ones in comparison with that from $R_b$~\cite{d2hdm}.
Combining the theoretical constraints with the experimental ones as done in Refs.~\cite{sk1,sk}, we can investigate how  the masses of Higgs bosons and mixing parameters can be constrained.

\section{Results and Discussion}
\label{sec:numresults}

\begin{figure}[tb]
    \includegraphics[scale=0.8]{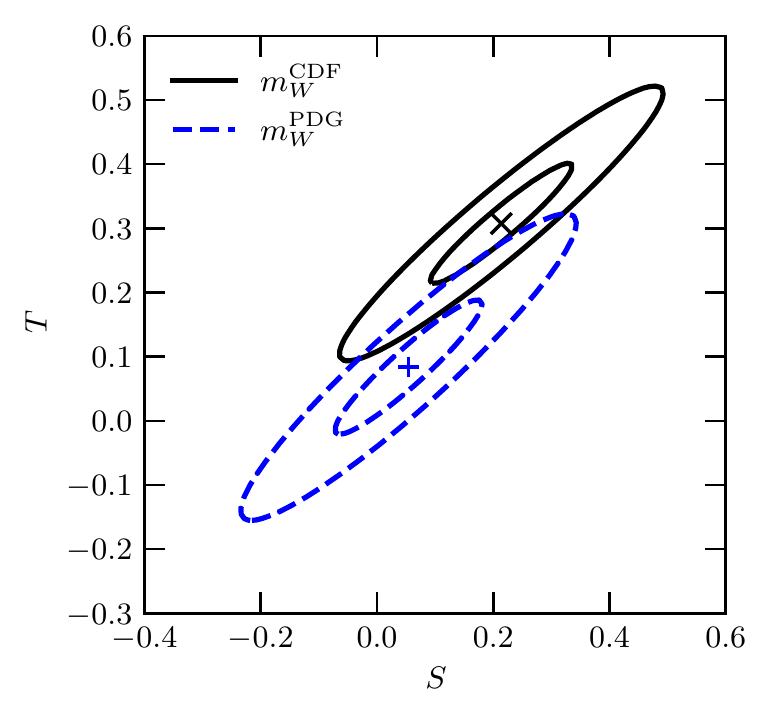}%
    \caption{\label{fig:tscontour}%
        Contours for the S and T oblique parameters obtained with Eqs.~\eqref{st-values-mw} and~\eqref{st-values}.
        The black solid contours correspond to the value obtained using the value recently announced by the CDF Collaboration,
        the blue dashed contours is obtained using the previous average $80.379\pm0.012$~GeV.
        The interior (exterior) contour corresponds to 1$\sigma$ (3$\sigma$).
        }
\end{figure}

In this section we present the analysis performed
and discuss relevant details of the results.
The contribution from the oblique parameter $U$ is expected to be negligible
compared to the contributions from $S$ and $T$.
Using Eqs.~\eqref{st-values} we obtain the limits cited in Eq.~\eqref{ew-parameter}.
The two-dimensional contours are shown in Fig.~\ref{fig:tscontour},
including the value obtained using the previous average
that is consistent with the SM\@.
In what follows we describe the rest of the components of our numerical analysis.

\subsection{Methodology}
\label{sec:numres:method}

We begin by enforcing the conditions of unitarity, perturbativity and
stability of the potential, by requiring that Eqs.~\eqref{stability} are respected.
These theoretical conditions are applied as hard cuts
to ensure that every sampled point is theoretically meaningful.
Considering that the oblique parameters, dominantly $S$ and $T$,
shift the value for $\sin^2\theta_W(m_Z)_{\overline{\text{MS}}}$,
we will require their values to be consistent
with both the recent measurement of the mass of the $W^\pm$ by the CDF collaboration
and with the current experimental average for the weak mixing angle.
To calculate the theoretical constraints and oblique parameters,
as well as other observables we employ \texttt{2HDMC}~\cite{Eriksson:2009ws}.
This tool is then interfaced with \texttt{HiggsBounds}~\cite{Bechtle:2020pkv}
and \texttt{HiggsSignals}~\cite{Bechtle:2020uwn}
to incorporate several constraints from LEP, Tevatron and LHC on the Higgs sector
and obtain a $\chi^2$ for the currently observed signals of the Higgs.
Finally, to calculate flavor physics observables,
that will be relevant mostly for the 2HDM-II as mentioned in Sec.~\ref{sec:thebounds},
we process our obtained data with \texttt{SuperIso}~\cite{Mahmoudi:2008tp}.
We use the Markov Chain Monte Carlo sampler \texttt{emcee}~\cite{ForemanMackey:2012ig} to explore the parameter space.
The free parameters of our model are given by 
the mass of the heavy Higgs, $m_H$,
the charged Higgs mass, $m_{H^\pm}$,
the mass of the pseudoscalar, $m_A$,
the mixing $\cos(\beta - \alpha)$,
the parameter $\tan\beta$
and the squared mass parameter $m_{12}^2$.
The limits of our parameter scan are given by:
\begin{align}
    M_H/1~\text{TeV}:&\quad  \text{type I: } [0.13, 1], \text{ type II: } [0.13, 1.9], \nonumber\\
    M_{H^\pm}/1~\text{TeV}:&\quad \text{type I: } [0.08, 1], \text{ type II: } [0.15, 1.9]\nonumber\\
    \label{eq:parameterranges}
    M_A/1~\text{TeV}:&\quad \text{type I: } [0.02, 1], \text{ type II: } [0.02, 1.9],\\
    \cos(\beta - \alpha):&\quad [-0.5, 0.5], \nonumber\\
    \tan\beta:&\quad [0.1,20], \nonumber\\
    m_{12}^2/1~\text{GeV}^2:&\quad [0, 500^2],\nonumber
\end{align}
where the larger upper bound for type II considers the higher lower bound on $M_{H^\pm}$
imposed by flavor physics as mentioned in Sec.~\ref{sec:thebounds}.

The recent measurement of the mass of the $W^\pm$
that we use to constrain our parameter space is given in Eq.~(\ref{cdf}).
%\begin{equation}
%    \label{eq:cdfwmass}
%       m_W = 80433.5\pm 9.4~\text{MeV}.
%\end{equation}
We will also use the average published by the PDG collaboration, given by $m_W^\text{PDG} = 80.379\pm 0.012$,
to compare the parameter space that leads to these two different measurements.
We fix the mass of the lighter $CP$-even Higgs to the current average central value of the SM Higgs $m_h = 125.10$~GeV~\cite{ParticleDataGroup:2020ssz}.

\subsection{Scenario 1: 2HDM-I}

\begin{figure}[tb]
        \includegraphics[scale=0.8]{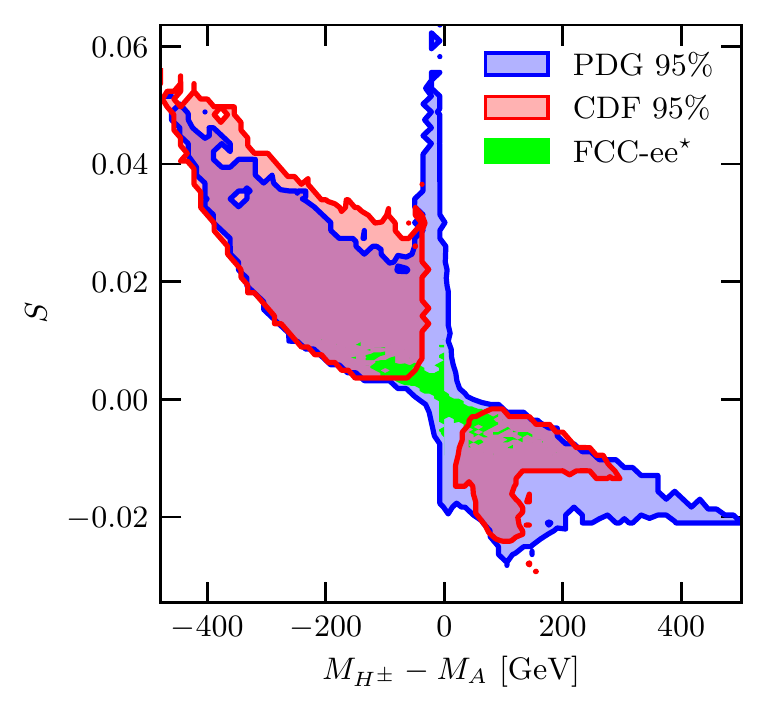}%
        \includegraphics[scale=0.8]{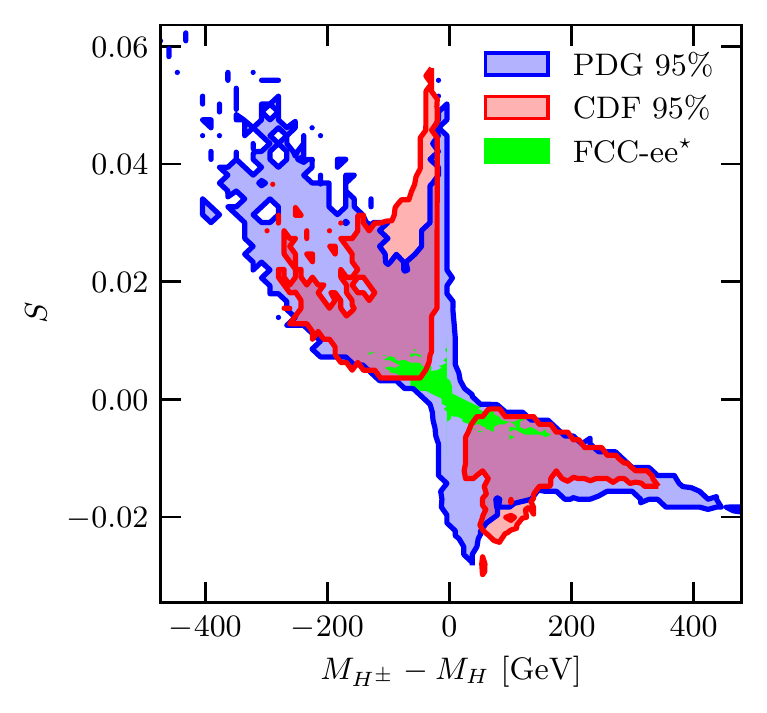}
        \includegraphics[scale=0.8]{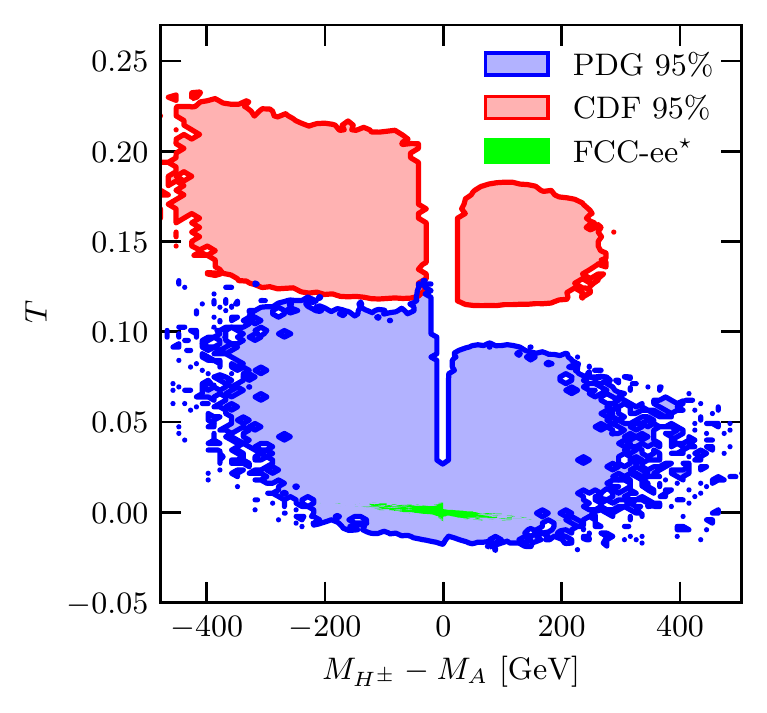}%
        \includegraphics[scale=0.8]{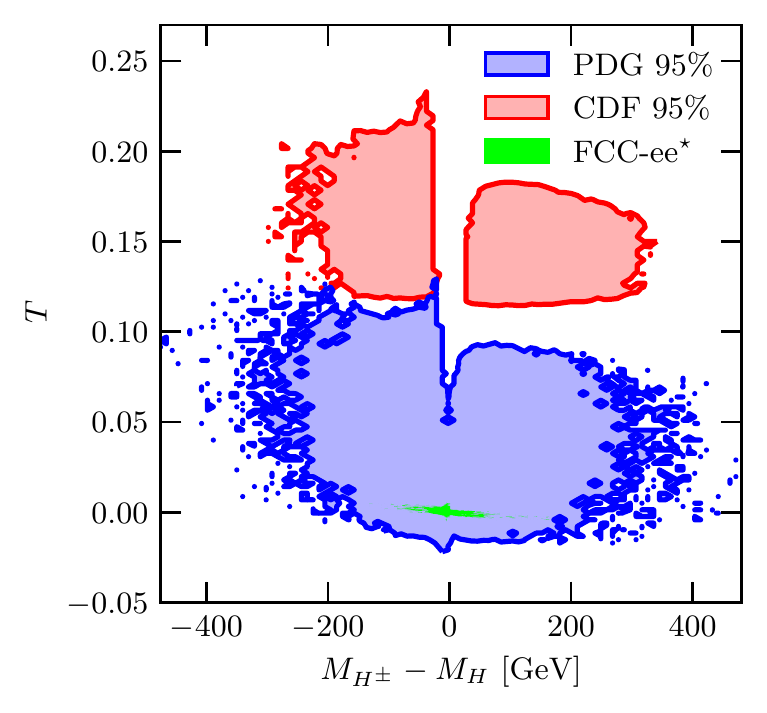}
        \includegraphics[scale=0.8]{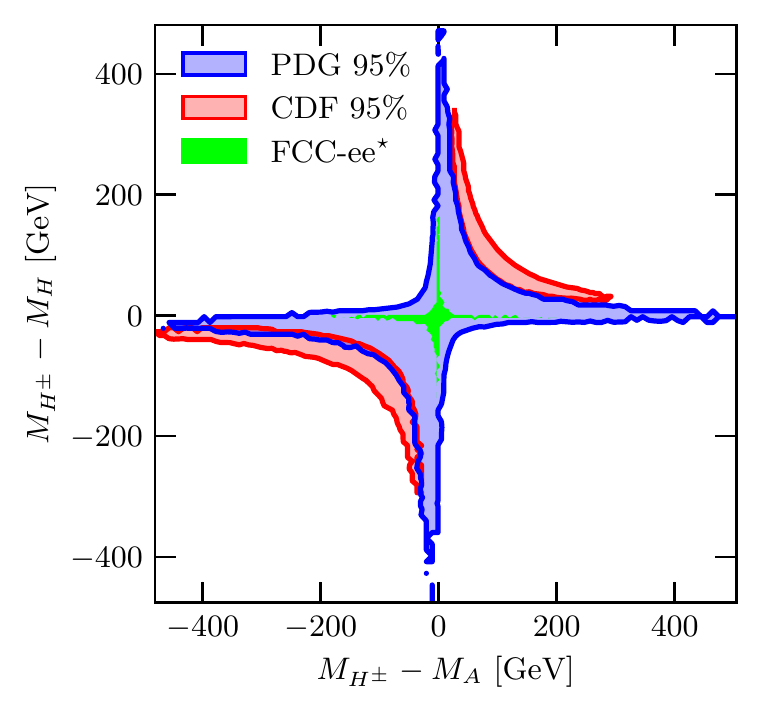}%
        \includegraphics[scale=0.8]{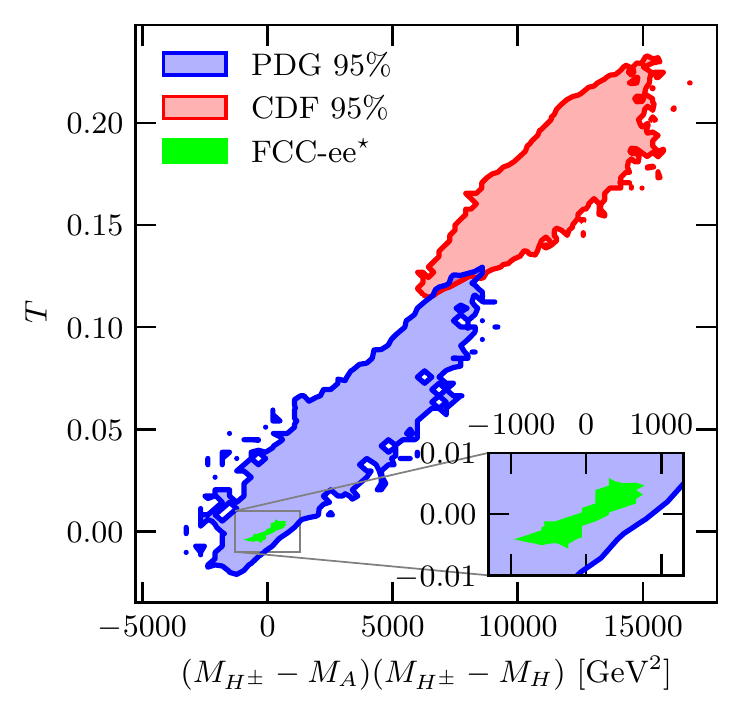}
    \caption{\label{fig:masssplittings}%
        Scenario 1, type I: Dependence of $S$ (upper) and $T$ (medium) parameters on the mass splittings
        $(M_{H^\pm} - M_A)$ (upper and medium-left) and $(M_{H^\pm} - M_H)$ (upper and medium-right).
        On the bottom, relation between mass splittings (left)
        and the dependence of $T$ on the product of these splittings (right).
        Regions consistent with CDF-II measurement of $m_W$ are shown in red,
        while regions in blue correspond to the average published by the PDG\@.
        The green region corresponds to the estimated sensitivity for FCC-ee~\cite{Blondel:2021ema}
        around the SM prediction
        ($^\star$see text for details).
        % All regions taken at 95\%~CL.
        }
\end{figure}

It is well known that the parameters $S$ and $T$
give the largest contribution to the shift in $m_W$.
From Eqs.~\eqref{T parameter} and~\eqref{S parameter} we can see that in the 2HDM
they are largely affected by mass differences between scalars.
Expectedly, we find that to predict the mass of the $W^\pm$ boson
in the range measured by the CDF Collaboration
we require sizable mass splittings.
Different mass splittings and their resulting values for $S$ and $T$
are shown in Fig.~\ref{fig:masssplittings}.
When comparing the contours at $95\%$ C.L. for the PDG average (blue) and the CDF measurement (red)
we see that the CDF measurement requires a higher value for $T$.
We also depict a green region for the projected sensitivity of FCC-ee~\cite{Blondel:2021ema} at 95\% C.L.
that will be explained in detail later.
In the case of the CDF measurement, for $S$ and $T$ we observe
that both splittings $M_{H^{\pm}} - M_A$ and $M_{H^\pm} - M_H$
are required to be simultaneously and always nonzero.
Closer inspection of Eqs.~\eqref{T parameter} and~\eqref{S parameter}
shows that if both splittings vanish simultaneously
then $S$ and $T$ would vanish as well.
This case results in $m_W$ as predicted by the SM
and is well inside the region using the PDG average value.
From the separation of the contours in the figures with $T$ axis,
we can infer that $T$ has the largest contribution to the shift in the mass of the $W^\pm$.
In the bottom panels of Fig.~\ref{fig:masssplittings}
we can see the relation between the two mass differences,
$M_{H^{\pm}} - M_A$ and $M_{H^\pm} - M_H$ (bottom left),
and that the allowed regions of their combination are almost separated for both cases of $m_W$ (bottom right).
In particular, we see that the regions follow hyperboliclike contours
that tend to be in regions where both mass differences have the same sign,
with the CDF measurement in regions with simultaneous larger mass differences.
This hyperbolic behavior hints to an effective dependence of $T$
on the product of mass differences,
which is demonstrated in the bottom-right panel of Fig.~\ref{fig:masssplittings}.

\begin{figure}[tb]
    \includegraphics[scale=0.8]{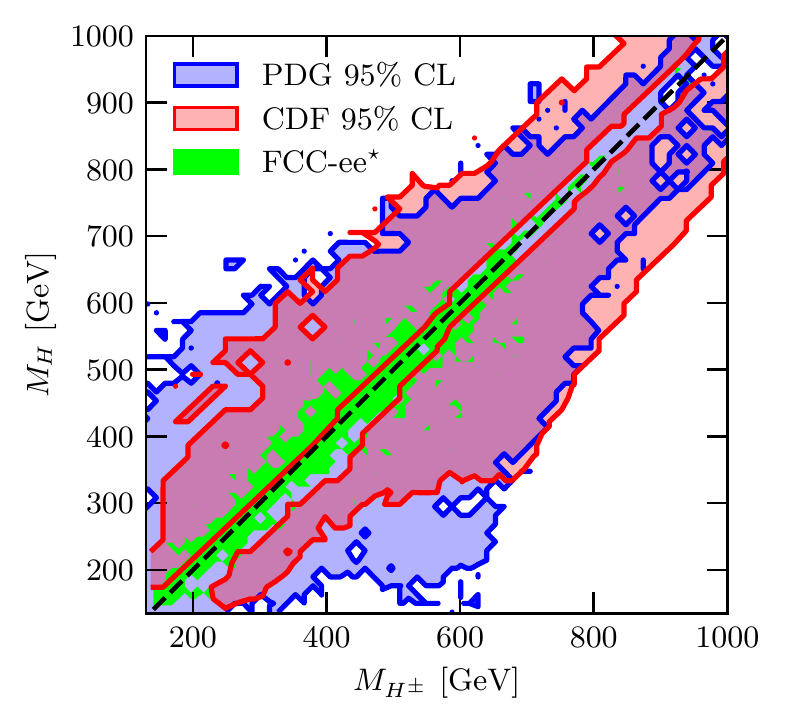}%
    \includegraphics[scale=0.8]{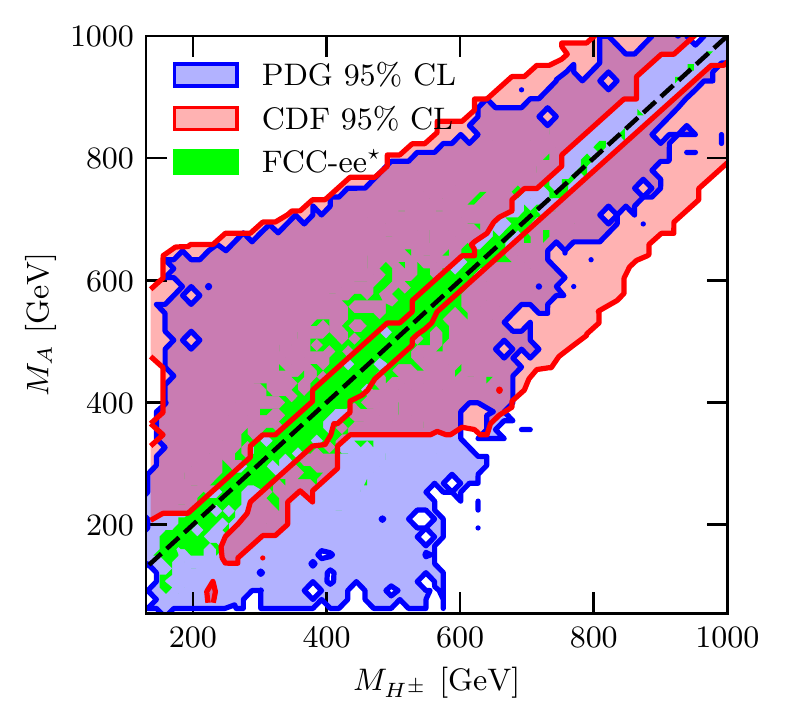}
    \caption{\label{fig:massranges}%
        Scenario 1, type I: Dependence of the scalars and pseudoscalar masses on each other.
        The colors are as in Fig.~\ref{fig:masssplittings}.
        The diagonal $x=y$ is represented by the dashed black line.
        }
\end{figure}

\begin{figure}[tb]
    \includegraphics[scale=0.8]{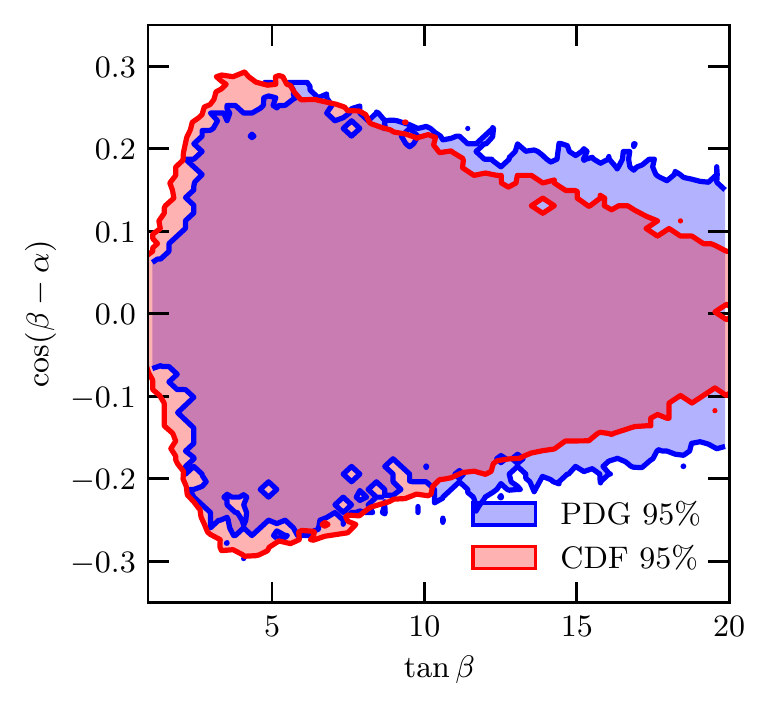}
    \caption{\label{fig:mixing}%
        Scenario 1, type I: Relationship between $\tan\beta$ and $\cos(\beta - \alpha)$.
        The colors are as in Fig.~\ref{fig:masssplittings} except for FCC-ee contour
        which has almost no effect for this combination of parameters.
        There is some reduction in $\cos(\beta - \alpha)$ range for large $\tan\beta$ due to data
        thinning that does not affect other results.
        }
\end{figure}

Since the size of the parameters $S$ and $T$
depends heavily on the splittings between scalars and pseudoscalar
we can expect their mass ranges to depend notably on the masses
of each other.
This is illustrated in Fig.~\ref{fig:massranges},
where the first notable feature
is how the diagonal for equal masses splits
the CDF region in two,
something that does not happen for the PDG average.
This is because of the need of nonzero mass differences as shown in Fig.~\ref{fig:masssplittings}.
The relationship between $\tan\beta$ and $\cos(\beta - \alpha)$,
the latter being the mixing between Higgses,
is show in Fig.~\ref{fig:mixing}.
In this case we can see that the range of $\cos(\beta - \alpha)$
is not obviously affected by the change in $W^\pm$ mass.

\begin{figure}[tb]
        \includegraphics[scale=0.8]{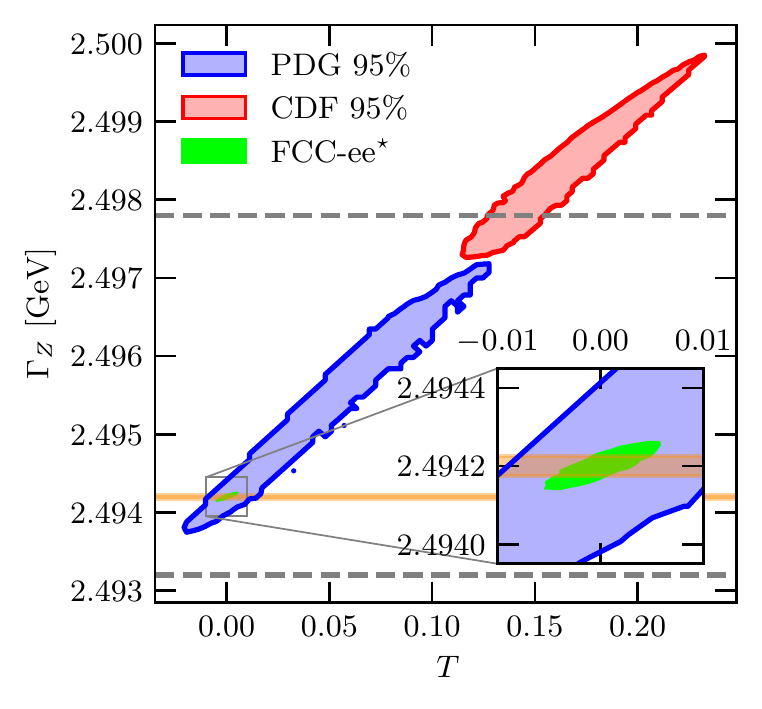}
    \caption{\label{fig:zdecayT}%
    Scenario 1, type I: Predicted decay width of the $Z$ boson, $\Gamma_Z$,
    for the allowed CDF and PDG regions.
    The dashed lines and orange band mark the 1$\sigma$ range for $\Gamma_Z$
    using the current average~\cite{ParticleDataGroup:2020ssz} and
    the FCC-ee projected sensitivity, respectively.
    The green region corresponds to the estimated sensitivity for FCC-ee
    around the SM prediction for several observables
    ($^\star$see text for details).
    }
\end{figure}

Up to this points, it is clear that the biggest difference between CDF and PDG regions
appears when projecting on the $T$ parameter.
The size of the $T$ parameter strongly affects
the prediction for the decay width of the $Z$~\cite{Burgess:1993mg,Ciuchini:2013pca}
(see also Sec.~10 of Ref.~\cite{ParticleDataGroup:2020ssz}),
which is explicitly demonstrated in Fig.~\ref{fig:zdecayT}.
Besides the decay width of the $Z$ boson,
both parameters, $S$ and $T$, can have effects
on other electroweak precision observables.
With this in mind, we estimate the future allowed region
using the projected error bars from the proposed
Future Circular Lepton Collider (FCC-ee)
as reported in Table 3 of Ref.~\cite{Blondel:2021ema}.
In particular, we use projected sensitivities for
measurements of $Z$ properties: decay width ($\Gamma_Z$),
the ratios of hadrons to leptons partial decay widths ($R^Z_\ell$),
the ratio of $b\overline{b}$ to hadrons partial decay widths ($R_b$)
and the hadronic cross section ($\sigma_\text{had}$),
as well as the projected error bar on the mass of the $W$.
We add statistical and systematic errors in quadrature
and assume that the central value will be consistent with the SM\@.
The resulting constraints on the allowed region is quite severe
as is clearly seen in the 95\% C.L. green regions in Figs.~\ref{fig:masssplittings} and~\ref{fig:massranges}.
In Fig.~\ref{fig:masssplittings} we can see that the sizes for $S$ and $T$
are reduced down to $\mathcal{O}(10^{-2})$ or less.
The effect on the mass splittings is more clear when we consider the combination
of them, as can be seen in the bottom panels of the same figure.
We can see that for FCC-ee, at least one of the two mass differences displayed
has to be close to zero resulting in a very small allowed region for their product.
This is reflected in Fig.~\ref{fig:massranges},
where we can find the FCC-ee consistent regions mostly around the dashed diagonal line.
Considering the results of this analysis,
if the 2HDM-I is responsible for the deviation in the mass of the $W$,
the FCC-ee should see further deviations in other observables,
most notably $\Gamma_Z$.
So, we can expect that the FCC-ee would decisively support or rule out
the CDF measurement.
To finalize this part, the FCC-ee improvement is not expected to have an important effect on $\tan\beta$
or $\cos(\beta - \alpha)$ and is, therefore, not shown in Fig.~\ref{fig:mixing}

\subsection{Scenario 1: 2HDM-II}

\begin{figure}[tb]
        \includegraphics[scale=0.8]{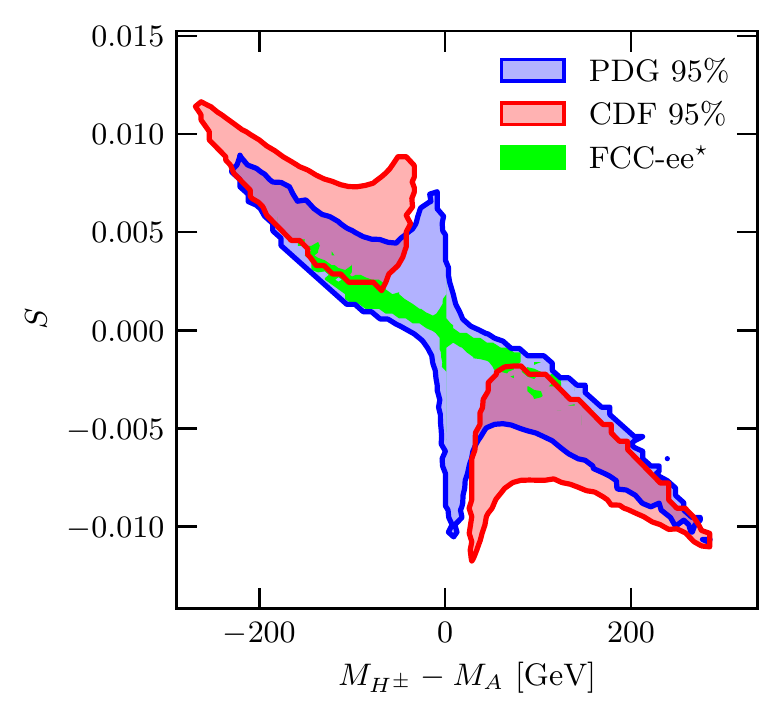}%
        \includegraphics[scale=0.8]{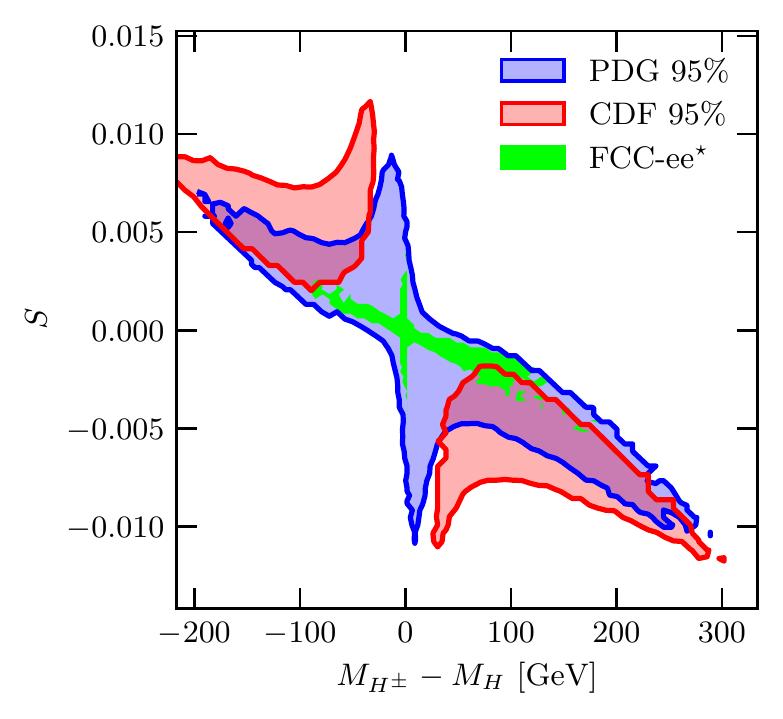}
        \includegraphics[scale=0.8]{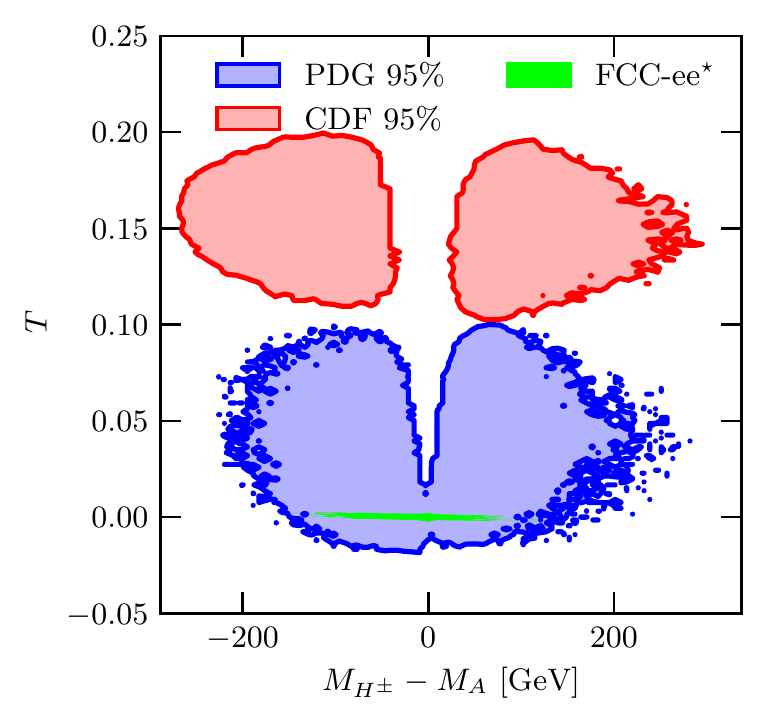}%
        \includegraphics[scale=0.8]{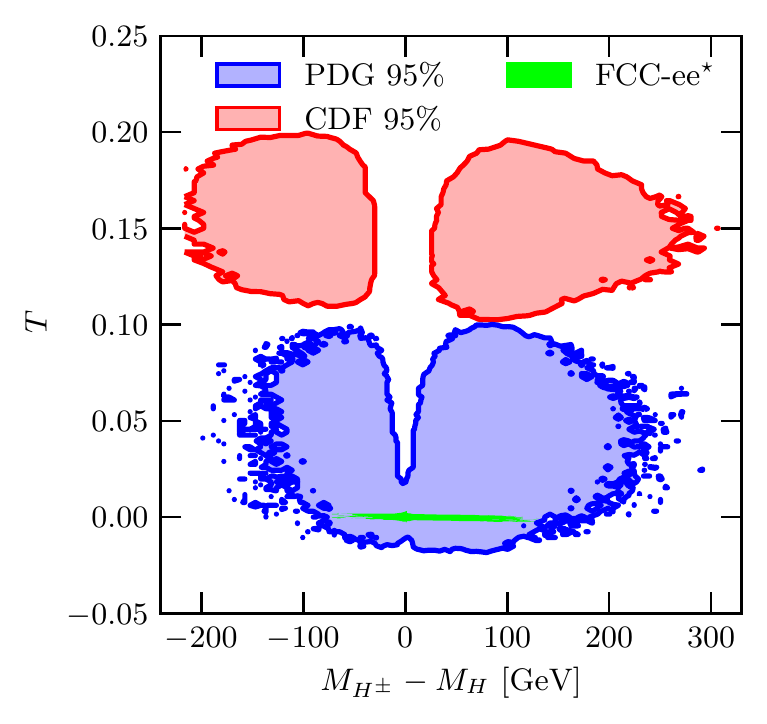}
        \includegraphics[scale=0.8]{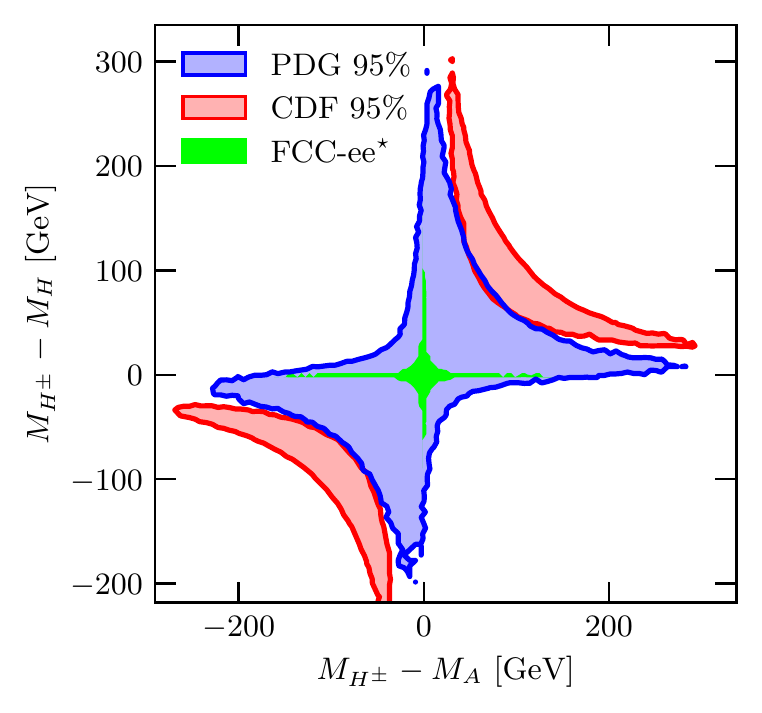}%
        \includegraphics[scale=0.8]{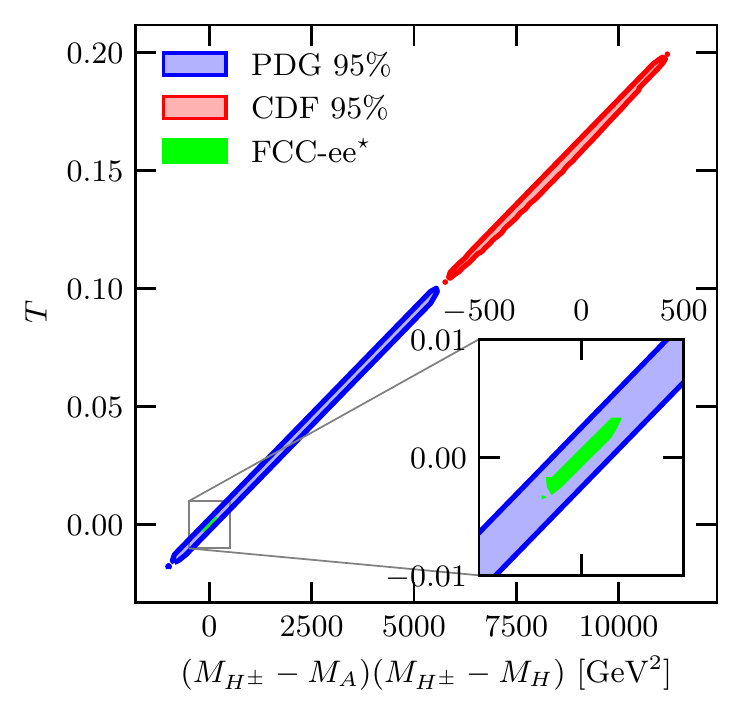}
    \caption{\label{fig:masssplittingsii}%
        Scenario 1, type II: Dependence of $S$ (upper) and $T$ (medium) parameters on the mass splittings
        $(M_{H^\pm} - M_A)$ (upper and medium left) and $(M_{H^\pm} - M_H)$ (upper and medium right).
        On the bottom, relation between mass splittings (left)
        and the dependence of $T$ on the product of these splittings (right).
        The colors are as in Fig.~\ref{fig:masssplittings}.
        }
\end{figure}

In the case of the 2HDM-II we see some of the same features
such as the need for non zero mass differences for $H^\pm$, $A$, and $H$.
As before, in Fig.~\ref{fig:masssplittingsii} we see a clear separation
between the values of $T$ that are required for the $W^\pm$ mass
measured by CDF and the PDG average.
However, we see that in this case the range of the difference $M_{H^\pm} - M_A$
is slightly smaller than it was for 2HDM-I.
For $S$, we find values much smaller ranging approximately between [$-0.01$,$0.01$]
and we see less overlap between both regions.
In the bottom panels of Fig.~\ref{fig:masssplittingsii},
we see again that the mass differences have a hyperbolic behavior,
as in the case of 2HDM-I, although with smaller ranges.
This results in a very narrow relationship
between the product of mass differences and $T$,
as can be seen on the bottom right panel.

\begin{figure}[tb]
        \includegraphics[scale=0.8]{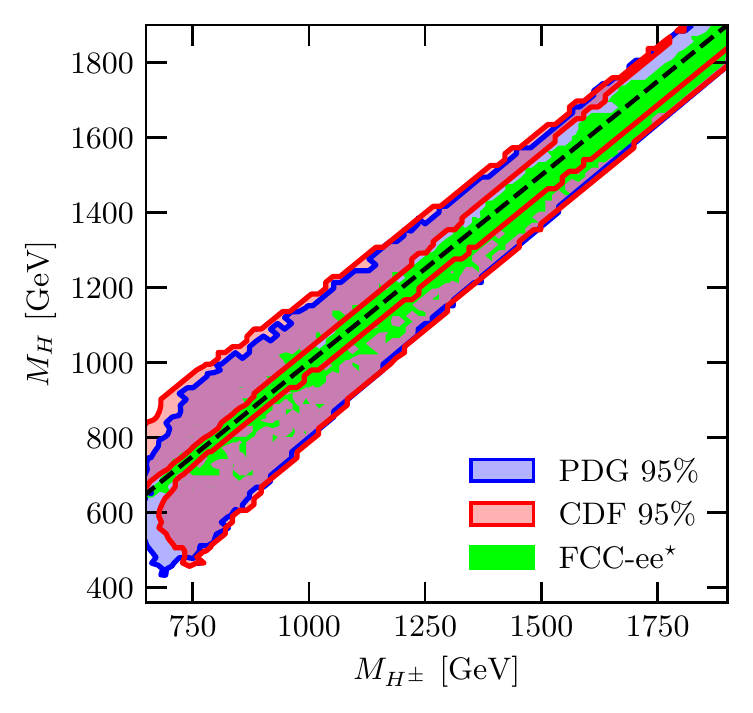}%
        \includegraphics[scale=0.8]{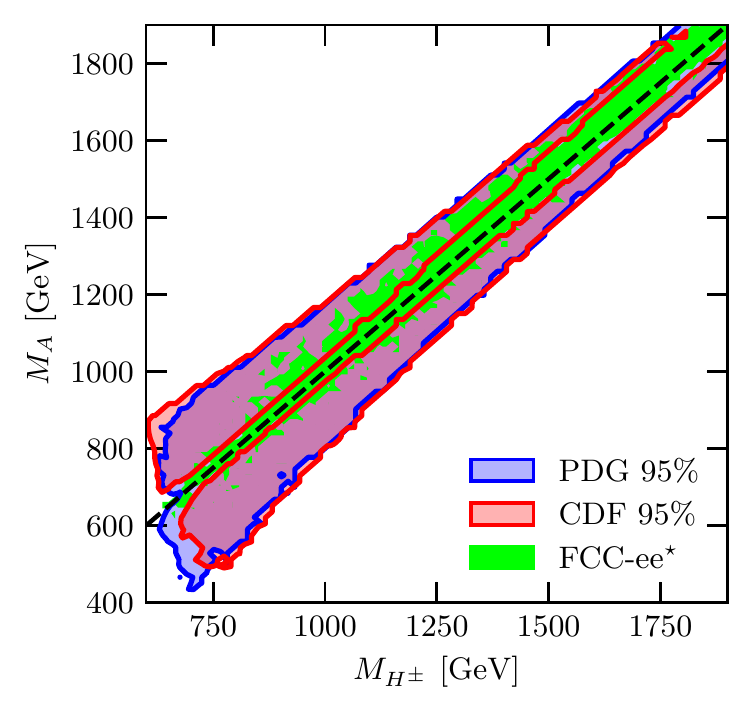}
    \caption{\label{fig:massrangesii}%
        Scenario 1, type II: Dependence of the scalars and pseudoscalar masses on each other.
        The colors are as in Fig.~\ref{fig:masssplittings}.
        The diagonal $x=y$ is represented by the dashed black line.
        }
\end{figure}

The most notable feature of 2HDM-II is the effects of the charged Higgs,
$H^\pm$, in flavor physics observables, most importantly in $\bar{B}\rightarrow X_s \gamma$,
resulting in a lower bound for $M_{H^\pm}$.
In Fig.~\ref{fig:massrangesii} we see the allowed regions projected in
the planes $M_{H^\pm}$-$M_H$ (left) and $M_{H^\pm}$-$M_A$ (right).
Again, we see that the equal masses diagonal splits the CDF regions in two sections.
In the same figure, both panes display a lower bound for $M_{H^\pm}$ that is taken to be around 650~GeV,
well inside the expected range
for the 95\% C.L. lower bound
found in Ref.~\cite{Misiak:2017bgg},
although the precise value could change with a proper, dedicated study on the
observables affected by the $H^\pm$.
Such study is out of the scope of this work.

\begin{figure}[tb]
        \includegraphics[scale=0.8]{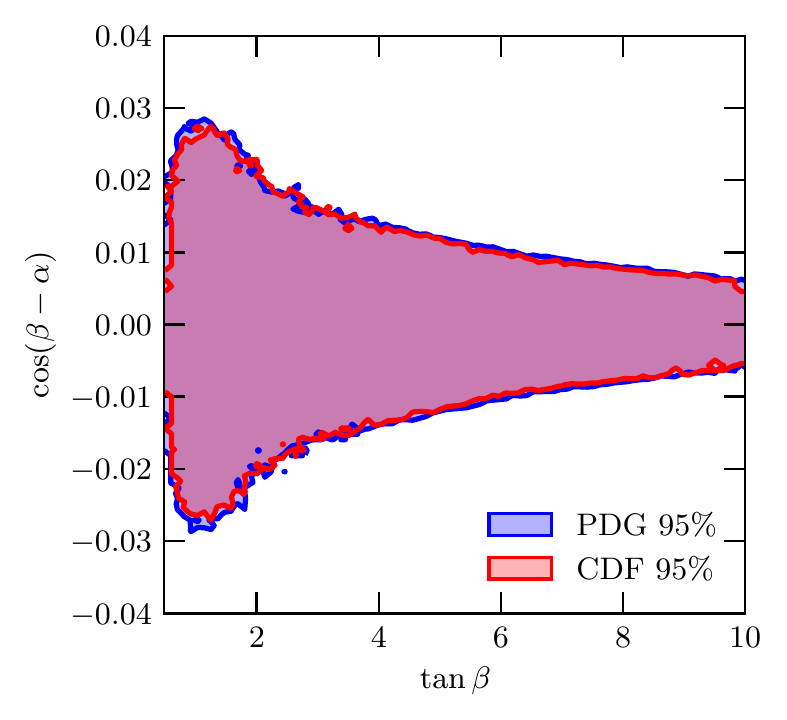}
    \caption{\label{fig:mixingii}%
        Scenario 1, type II: Relationship between $\tan\beta$ and $\cos(\beta - \alpha)$.
        The colors are as in Fig.~\ref{fig:masssplittings} except for the FCC-ee contour,
        which has almost no effect for this combination of parameters.
        }
\end{figure}

We show the relationship between $\tan\beta$ and $\cos(\beta - \alpha)$
in Fig.~\ref{fig:mixingii}.
Similar to type I, there is no considerable change in the allowed contours
from using the two different results for $m_W$.
However, we note that in this case the resulting range for $\cos(\beta - \alpha)$
is approximately 10 times smaller that it was for type I, reflecting that  the decoupling limit $\sin(\beta-\alpha) \sim 1$ is preferred.

\begin{figure}[tb]
        \includegraphics[scale=0.8]{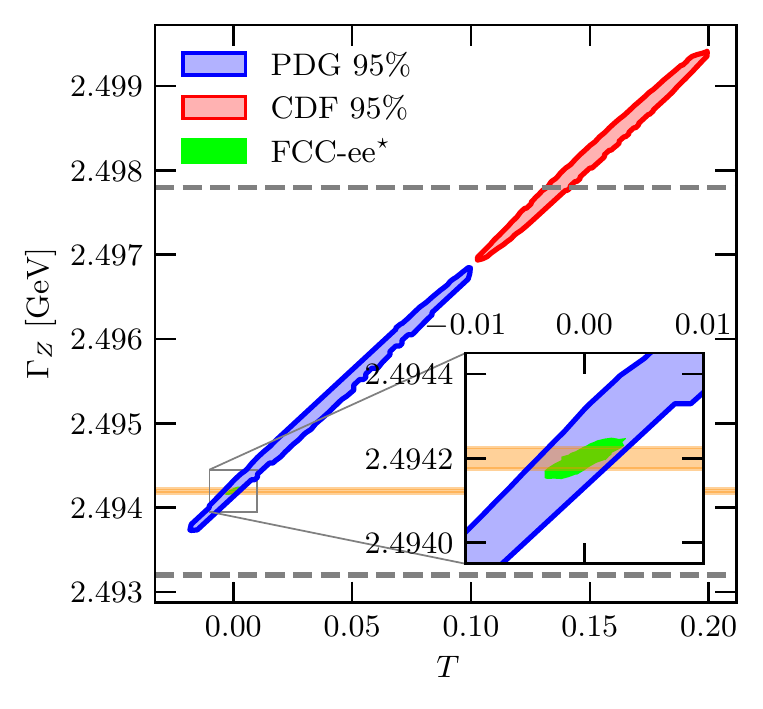}
    \caption{\label{fig:zdecayTII}%
    Predicted decay width of the $Z$ boson, $\Gamma_Z$,
    for the allowed CDF and PDG regions.
    The dashed lines and orange band mark the 1$\sigma$ range for $\Gamma_Z$
    using the current average~\cite{ParticleDataGroup:2020ssz} and
    the FCC-ee projected sensitivity, respectively.
    The green region uses the estimated sensitivity for FCC-ee
    around the SM prediction for several observables
    ($^\star$see text for details).
    }
\end{figure}

In this case we also apply the same analysis as in the previous part,
using the projected sensitivity by FCC-ee.
Expectedly, the results are very similar to the ones described for 2HDM-I above,
with $S$ and $T$ reduced to $\mathcal{O}(10^{-2})$ sizes or less
and at least one mass difference forced to be much closer to zero than the other.
Due to the narrower ranges for the mass differences, that also result in a much narrower $S$,
the predicted regions at $95\%$~C.L. for the decay width of the $Z$ in Fig.~\ref{fig:zdecayTII}
are much thinner than in the case of the 2HDM-I as can be seen in Fig.~\ref{fig:zdecayT}. 
However, the same conclusion can be drawn about FCC-ee
being able to observe deviations in other observables, such as $\Gamma_Z$.
Again, the FCC-ee projected sensitivity
does not significantly affect $\tan\beta$ and $\cos(\beta - \alpha)$
and we do not show its region in Fig.~\ref{fig:mixingii}.

\subsection{Scenario 2: 2HDM-I}
\label{sec:scen2THDMI}

\begin{figure}[tb]
        \includegraphics[scale=0.8]{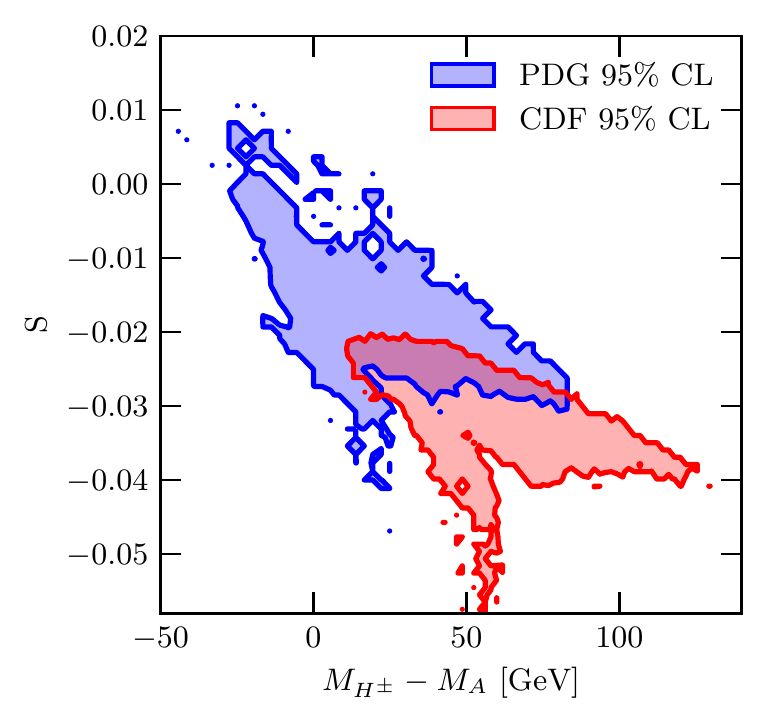}%
        \includegraphics[scale=0.8]{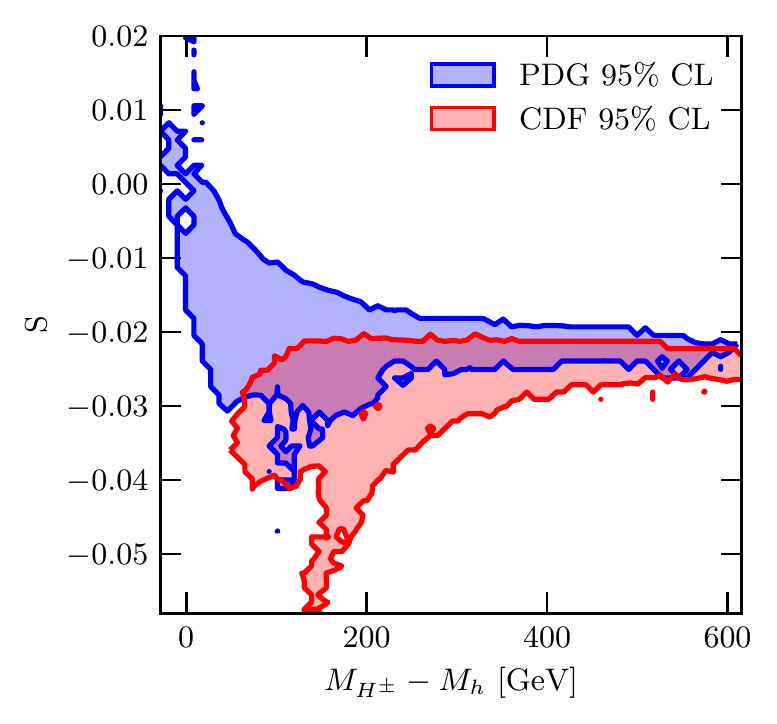}
        \includegraphics[scale=0.8]{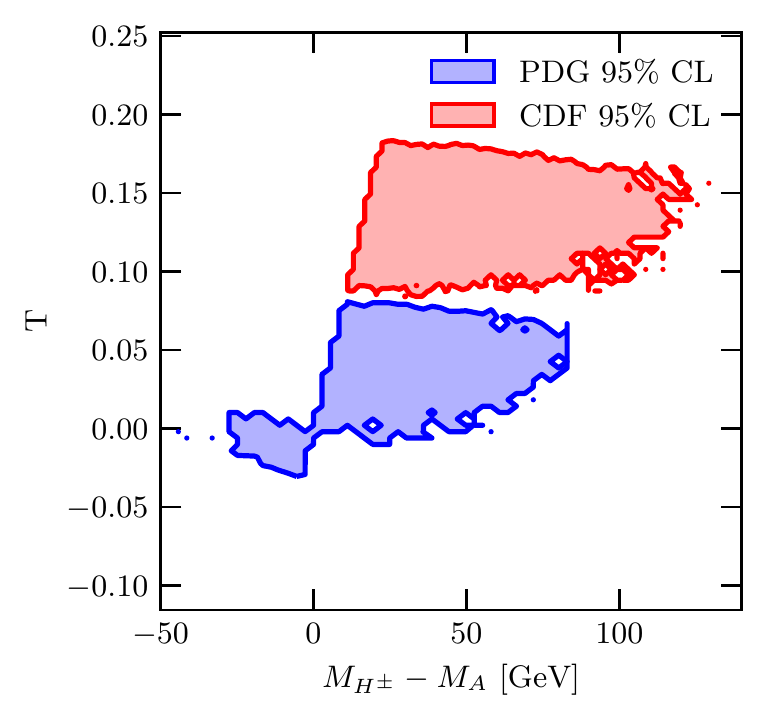}%
        \includegraphics[scale=0.8]{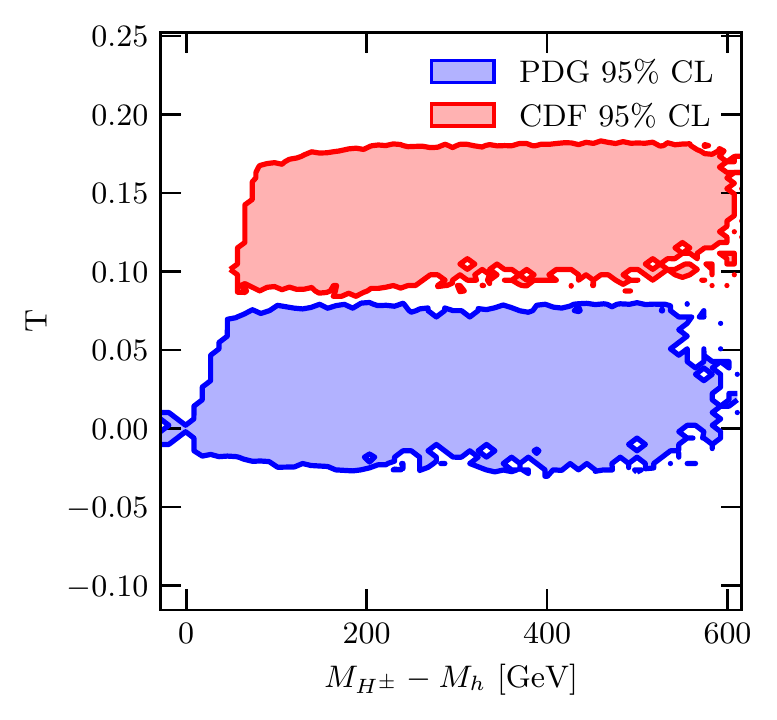}
    \caption{\label{fig:masssplittingsi2}%
        Scenario 2, type I: Dependence of the $S$ (upper) and $T$ (lower) parameters on the mass splittings
        between $H^\pm$--$A$ (left) and $H^\pm$--$H$ (right).
        For the red contour, tension with measured $\sin^2\theta_W(m_Z)_{\overline{\rm MS}}$ has been ignored
        to show the regions that reproduces the CDF measurement of $m_W$.
        }
\end{figure}

Scenario 2, where the mass of the lighter $CP$-even neutral scalar
is chosen as $M_h < 125.1$~GeV,
presents several complications due to the loss of freedom compared
to scenario 1.
First of all, considering that the mass differences
required to predict the CDF measured $m_W$
belong in the ranges of $\mathcal{O}(10)$~GeV to a few $\mathcal{O}(190)$~GeV
we can expect the mass difference $M_{H^\pm} - M_h$ predominantly positive.
From what we saw in the results of previous cases
we can say that this will limit the values of $S$ to be mostly negative.
This will heavily affect our allowed ranges,
since, from Eq.~\eqref{ew-parameter}, we know that $S$
is preferred positive.
In fact, from our analysis, we find that this scenario results
in a value for $\sin^2\theta_W(m_Z)_{\overline{\rm MS}}$
that is just marginally inside its $3\sigma$ region.
This is confirmed in Fig.~\ref{fig:masssplittingsi2},
where we can see in the upper two panes that CDF measurement of $m_W$
prefers $S$ below $-0.02$.
Note that in Fig.~\ref{fig:masssplittingsi2} the tension with the measured $\sin^2\theta_W(m_Z)_{\overline{\rm MS}}$
has been ignored for the red contour to display the region that predicts $m_W^{\rm CDF}$.
Also, we see that the preferred range for $M_{H^\pm} - M_h$ is almost entirely positive
and reaches up to $\sim 600$~GeV.
In the case of $T$, this results in one connected region corresponding to positive mass differences
in contrast to the scenario 1 where $T$ had two disconnected regions
for positive and negative mass differences.
From scenario 1 we learned
that the preferred values of $S$ and $T$ happened simultaneously
for $M_{H^\pm} - M_A$ and $M_{H^\pm} - M_A$ negative,
therefore, we can conclude that 2HDM-I in scenario 2 is disfavored
by requiring a light $CP$-even neutral state.
Considering this result, in this scenario we do not apply the analysis on future prospects for FCC-ee.

\subsection{Scenario 2: Brief comment on 2HDM-II}

From Sec.~\ref{sec:scen2THDMI} we learned that requiring a $CP$-even neutral state lighter than 125.1~GeV
resulted in less freedom and more constrained parameter space.
In particular, having the correct size for $M_{H^\pm} - M_h$ in this case results
in a lighter $M_{H^\pm}$ since the required mass differences are of a few 100~GeV at most.
This is particularly problematic in 2HDM-II,
where $\bar{B}\rightarrow X_s \gamma$ strongly constraints
light $M_{H^\pm}$.
We find that the mass $M_{H^\pm}$ required to predict $m_W^{\rm CDF}$
is light enough to be in tension with flavor physics measurements,
since, as can be seen in Fig.~\ref{fig:massrangesii},
the charged Higgs is constrained to be above $\sim 600$~GeV.

\section{Conclusion}
\label{sec:conclusion}

In this work we attempted to constrain the 2HDM parameter space
based on the new measurement of a 7$\sigma$ deviation from the SM
for the mass of the $W^\pm$ boson, $m_W$.
Interestingly, we found that by combining
the relationship of $m_W$ and $\sin^2(m_Z)_{\rm MS}$
with the $S$ and $T$ parameters we can constrain those parameters
at a level consistent with more complete global fits~\cite{Lu:2022bgw}.
We demonstrated that using this constraint,
together with the usual theoretical conditions
and several observations from LEP, Tevatron and LHC,
there is a set of parameters that is compatible with the new
measurement and, therefore, the 2HDM could successfully survive
if the deviation on $m_W$ is confirmed in the future.
In particular, we found that the scenario where the Higgs found at the LHC
having 125.1~GeV is identified as the light $CP$-even neutral scalar
is more favored compared to the scenario where it corresponds to the heavier state.
We show how important the mass splittings are
to shift the $W^\pm$ mass from its value predicted in the SM,
mostly via the contribution from the oblique parameter $T$.
We found clear differences between the requirements of the parameter space
depending on the predicted $W^\pm$ mass.
Most notably, the required value of $T$ takes nonoverlapping regions
when one considers the two options of the CDF measurement and the PDG average.
On the side of future prospects, we showed how an improved measurement
of $Z$ pole observables at the FCC-ee
should clearly distinguish 2HDM hints on deviations to the SM\@.
In our case, we found that if the 2HDM explains the newly measured $m^\text{CDF}_W$
FCC-ee should observe a clear deviation from the SM expectation,
particularly for the decay width of the $Z$.
Indeed, future experimental observations may bring more exciting
clues about where to focus theoretical efforts.

\end{document}